\documentclass[paper,copyright,creativecommons]{eptcs}
\providecommand{\event}{SYNT 2014} % Name of the event you are submitting to
\usepackage{breakurl}             % Not needed if you use pdflatex only.

\usepackage{graphicx}

\title{Synthesis of a simple self-stabilizing system}
\author{Adri{\`a} Gasc{\'o}n
\institute{SRI International\\
 California, USA}
\email{adriagascon@gmail.com}
\and
Ashish Tiwari
\institute{SRI International\\
California, USA}
\email{tiwari@csl.sri.com}
}
\def\titlerunning{Synthesis of a simple self-stabilizing system}
\def\authorrunning{Adri{\`a} Gasc{\'o}n, Ashish Tiwari}
\begin{document}
\maketitle

\sloppy

\newcommand\comment[1]{}

\begin{abstract}

%Designing correct, especially fault tolerant, distributed algorithms is an
%extremelly challenging task.  
%The main source of difficulty comes from the
%complex variety of scenarios
%to be handled that result from 
%concurrent execution of various agents.
%Moreover, design and implementation errors
%in this area are often difficult to identify, reproduce, and hence
%fix, since they may manifest only under very complex
%circumstances. 

With the increasing importance of distributed systems as 
a computing paradigm, a 
systematic approach to their design %of distributed systems
% that can overcome such inherent difficulties 
is needed.
Although the area of formal verification has made enormous
advances towards this goal, % in the last decade, 
the resulting functionalities are limited to detecting problems in a particular design.
By means of a classical example, 
we illustrate a simple template-based approach to computer-aided design of
distributed systems based on leveraging
the well-known technique of bounded model checking
to the synthesis setting.
%This projects pushes those capabilities forward by 
%providing the designer with synthesis tools that can
%actively guide the design process, that is, do computer-aided design.

%Our approach to computer-aided design of distributed synthesis
%is inspired by a lesson recently learnt by the program synthesis
%community: % that already lead to impressive results: 
%although the problem of automatically generating programs
%from their specification is inherently difficult,
%a design process based on the  
%interaction between a human a computer, where the human provides general insight
%and the computer works out the details, can be very effective.

\end{abstract}

\section{Introduction}

Consider a situation where a developer is trying to
design and program a multi-agent distributed system
to perform a certain task.
The agents could be 
robots communicating with each other,
sensors in a sensor network, 
processes in a multi-core machine,
or processors connected to a bus in an avionics system.
The task could involve achieving consensus,  
getting mutually exclusive access to some shared resource, 
computing some function of some sensed data, or
something similar.
The developer knows the underlying topology of the
communication network, the (synchronous or
asynchronous) communication model, as well as the nature of faults
the network and the agents themselves can manifest.
Does there exist a known algorithm that the developer 
could use in this particular scenario?

One path forward for the developer would be to study the literature on
distributed algorithms~\cite{lynch-book}. % to find out.
%The literature 
It contains several impossibility results,
as well as positive results and algorithms for 
several distributed problems,
but all those results are accompanied with an applicability
criterion.  Does the condition for applicability hold 
for the developer's particular situation?
Suppose that an impossibility
result is applicable.  Is there
a workaround if certain system requirements are changed?  
Suppose the developer finds an algorithm that appears
to be meaningful in his context, but the applicability
criteria does not match.  Can the discovered algorithm 
be massaged for his particular context?

%The story about the developer might sound fictitious.
%The truth is that it happened to us.  We were the 
%developers who were going through the above
%process, spending a large amount of time to answer
%each question, and then generating more new and harder
%questions from those answers. And all this in a seemingly infinite 
%loop of iteratively deeper investigation.

All the questions raised above may be difficult to answer.
The reason is that all intuitions about what 
works start to fail for distributed algorithms, and
more so in the presence of faults.
Reasoning about the correctness
of an algorithm in the presence of faults 
is not only difficult, but also a
surprisingly delicate task.
The following quote, taken from the seminal
paper that
introduced the well-known Byzantine Generals Problem~\cite{byzantine-generals}, 
talks about the correctness of an informally presented argument:

\begin{quote}
{\em{
``This argument may appear convincing, but we strongly advise the reader to be
very suspicious of such nonrigorous reasoning. Although this result is indeed
correct, we have seen equally plausible `proofs' of invalid results.
We know of no area in computer science or mathematics in which informal reasoning
is more likely to lead to errors than in the study of this type of algorithm.''
}}
\end{quote}

For these reasons, applying formal verification
to distributed algorithms, as well as their 
fault-tolerant variants, has drawn considerable attention.
In fact, several mechanized correctness proofs exist for 
some classical distributed 
algorithms~\cite{Shankar, RushbyVonHenke, rushbyLincoln, ShankarRushbyVonHenkeOwre}, and such 
mechanization even
lead to the detection of flaws in published results in some cases~\cite{RushbyVonHenke, rushbyLincoln}.
In constrast to these examples of the application of verification technology,
where the goal was to formally verify the correctness of a given algorithm, 
we are interested in {\em{using formal methods to guide the human in the design process}},
instead of just to verify its result. 

%There is a clear need for some {\em{computer-aided support}}
%to help explore the answer to the questions raised above.
%%There is a clear need for a tool that can help developers
%%answer (at least some of) the questions  raised above.
%Formal verification tools are helpful, but only moderately so,
%since one has to guess and check iteratively using a 
%verification tool.  To really expedite the exploration process,
%there is a need for {\em{formal synthesis}} tools.

Recently, there has been lots of excitement and progress
in automatically synthesizing programs that satisfy some
given requirement.
This development was triggered by the observation that
program synthesis starts becoming feasible if we start
from a program sketch -- rather than from a clean slate -- 
and synthesize a program by completing the sketch so that
it matches the given specification.
This synthesis process has been effectively demonstrated 
for imperative programs~\cite{DBLP:conf/pldi/Solar-LezamaRBE05, DBLP:conf/asplos/Solar-LezamaTBSS06}.  

The main reason why distributed algorithms are specially suitable for a
computer-aided design methodology is that the solutions to the 
kind of problems mentioned above are usually short and easy to describe,
while their correctness (or impossibility) proofs can be very involved.
This situation is ideal for synthesis tools, since their complexity
is roughly the product of the size of the design space and the
verification (checking) complexity.  Consequently, most synthesis tools
need either the design space to be small or the verification (checking)
effort to be minimal.  
%Moreover, using automation to help in designing the algorithm
%is more appealing since it can result in new insights, whereas
%using automation for verification is slightly less revealing 
%(at least to the designer who created the algorithm).

However, sketches for distributed algorithms can not be written in 
imperative languages.  A much richer language 
is needed (see~\cite{DBLP:conf/pldi/UdupaRDMMA13} for related
recent work on
synthesis distributed protocols).
The input language of formal verification tools, such as
the SAL language, is a great option.  It provides a very
rich set of constructs 
-- nondeterminism, synchronous and asynchronous
composition, parametric module specifications, module
instantiations, rich datatypes and rich expression language --
that are needed for modeling the execution of distributed algorithms 
in presence of faults.
However, the formal verification tools that run on SAL models, 
such as the SAL model checker, are just verification
tools and hence they do not perform synthesis.  If parts of the 
modeled system
are not known, they can not help complete the algorithm in any way;
though they can verify a manually completed sketch.

In this paper, we present 
computational techniques 
that can aid a human in exploring the design space of algorithms;
that is, the field of {\em{computer-aided synthesis}},
with a focus in problems arising in 
{\em{distributed systems}}. 
Our proposed approach is based on using {\em{synthesis-versions of popular
formal verification techniques}}.  
A general view of our approach 
to build computer-aided synthesis technology 
is shown in Figure~\ref{fig-approach}.
Just as SAT-based bounded model checking turns a verification 
problem into a search problem (over a large, but finite, search space),
QBF-based {\em{bounded model synthesis}} turns a synthesis problem into a
large, but finite, one-step $\forall\exists$ game that can be 
solved using a QBF solver.  Similarly, verification by 
$k$-induction can be lifted to $k$-inductive synthesis.
There is a similar correspondence
between the infinite state space versions of these techniques.
In the present paper we focus on bounded model
synthesis.
Our approach is enabled
by some impressive recent progress in the field of QBF solving and 
$\exists\forall$ SMT solving~\cite{DBLP:conf/sat/RanjanTM04,DBLP:conf/aspdac/YuM05,DBLP:journals/jsat/LonsingB10,DBLP:conf/cp/OlivoE11,DBLP:conf/sat/NiemetzPLSB12,DBLP:conf/sat/JanotaKMC12,DBLP:conf/lpar/JanotaGM13,DBLP:journals/corr/ChengSRB13}.

\begin{figure}[t]
\begin{tabular}{||l|ccccc||}
\hline
Task & Inputs & & Technique & & Backend solver
\\ \hline \hline
Verification: & Model, Property & $\longrightarrow$ & {\em{bmc, k-ind}} & $\longrightarrow$ & 
 SAT formula
 \\ \hline
 Synthesis: & Partial Model, Property & $\longrightarrow$ & {\em{bms, k-inds}} & $\longrightarrow$ & $\forall\exists$ QBF formula
 \\ \hline
Verification: & Model, Property & $\longrightarrow$ & {\em{inf-bmc, k-ind}} & $\longrightarrow$ & 
 SMT formula
 \\ \hline
 Synthesis: & Partial Model, Property & $\longrightarrow$ & {\em{inf-bms, k-inds}} & $\longrightarrow$ & $\forall\exists$ SMT formula
 \\ \hline
\end{tabular}
\caption{{\small{Lifting some verification techniques to synthesis: Bounded model checking (bmc) and $k$-induction (k-ind) are generalized to bounded model synthesis (bms) and $k$-inductive synthesis (k-inds); and similarly, for the infinite state space versions of these techniques.}}}\label{fig-approach}
\end{figure}

More concretely, in this paper we focus on leveraging the
technique of bounded model checking to the template-based synthesis setting. 
Our templates are written in the SAL language, which is, as
commented above, a suitable formalism to describe a
distributed system.
We also take advantage of the SAL model checker to 
contruct a 2QBF formula that is afterwards
sent to an off-the-shelf QBF solver.
Our work indicates that the synthesis-extension of bounded model checking
can be used to obtain surprising new 
algorithms, show non-existence of algorithms for
certain classes of problems,  and generate useful variants of known
algorithms.

The rest of this paper is organized as follows.
In the next section we present our running example,
a problem inspired by Dijkstra's paper~\cite{DBLP:journals/cacm/Dijkstra74}.
In Section~\ref{section:synthesisFG} we present our synthesis
methodology and describe an oportunistic implementation using the SAL model checker.
In Section~\ref{dec:running-example} we describe our experience
applying our approach to the running example,
with references to all the SAL models implemented along the way.
Finally, in Section~\ref{sec:discussion} we provide some discussion
and directions for further work.

\section{Running example: Reaching Mutual Exclusion}\label{sec:running-example}
In this section, we describe a simple example inspired
by Dijkstra's paper~\cite{DBLP:journals/cacm/Dijkstra74}, which is a remarkable milestone 
in the study of fault tolerance.
The example has the property that its solution is simple to describe, 
yet difficult to verify.
Subsequently, we will use the same example to 
present our approach to computer-aided
design of distributed algorithms. 
For another more complex example on fault-tolerant 
consensus, which also provides
a proof of concept for our approach, 
the reader is refered to~\cite{DBLP:conf/nfm/GasconT14}.

Consider a system with four machines $m_0, m_1, m_2, m_3$.
Each machine $m_i$ has two Boolean state variables $A$ and $B$.
The $4$ machines are arranged in a ring topology in which every
machine has read access to the state variables 
of its right and left neighbors; that is,
machine $m_i$ has read/write permission on 
its own state variables $A$ and $B$ and read permission on 
the state variables of machine $m_{(i+1){\tt mod}{4}}$, which we denote as 
$A_R$ and $B_R$, and the
the state variables of machine $m_{(i-1){\tt mod}{4}}$, which we denote as 
$A_L$, $B_L$.
Each machine $m_i$ updates its state 
according to a finite set of rules
$R_i$
of the form
$$ \mbox{{\sc if} privilege {\sc then} make move {\sc endif}} $$
where privilege is a Boolean condition on the state variables of 
the machine and its neighbors,
and a move is an update to the values of $A$ and $B$.
We say that a rule is {\em enabled} at some step if 
its privilege evaluates to true in that step
and its corresponding move {\em{changes}} the current state.
At each step, a rule is arbitrarily selected 
from the set of enabled rules 
and executed.
We say that the system is in a {\em legitimate} state if 
{\em{exactly one}}
rule in $\bigcup R_i$ is enabled.
The problem is to find rules for each machine $m_0,\ldots,m_3$ such that:
\begin{itemize}
\item[(a)] At least one rule will always be enabled and
    the system is {\em guaranteed} to reach a legitimate state {\em regardless
    of its initial conditions} in a {\em finite} number of steps.
\item[(b)] In each legitimate state, each possible move will
    bring the system again into a legitimate state.
\end{itemize}

Intuitively, the initial state is arbitrary and multiple machines can 
make a move, but eventually we want the machines to get a 
{\em{mutually exclusive access}} to make a move.

It is not at all obvious how to design local rules that
will achieve convergence towards states satisfying (a) and (b). 
Note that 
the source of difficulty is that the initial state, as well as
the subsequent moves of the system, are all picked nondeterministically.
Another source of difficulty that we will consider later is 
requiring {\em fairness}, i.e. for every pair of machines $m_i, m_j$, 
there is a sequence of steps of the system going from a legitimate state
where a rule of $m_i$ is enabled to a legitimate state
were a rule of $m_j$ is enabled. Note that there might be other
reasonable definitions
of fairness.

Some questions that may arise in the process of designing an
algorithm for this problem might be:
How many rules do we need? 
Is it useful to restrict the states space by fixing 
some variables to have a certain value?
Is there a solution where
all machines have the same set of rules?
As stated by Dijkstra in the original paper, the
discovery that the answer to the third question is ``no''
was crucial to obtaining an algorithm.

With our proposed approach,
the third question can be automatically answered for a 
fixed (but reasonably big)
number of rules of considerable complexity
and for a fixed choice for the number of steps to achieving convergence.

Before going into the details of 
our approach to template-based synthesis of 
distributed systems in the next section, 
let us present a possible solution to our problem.
We encourage the reader to think about the problem
at this point. 

In this solution,
$B_1$ is fixed to have value $false$, $B_4$ is fixed to have value $true$,
and the set of rules for each of the four machines is defined as follows:
\begin{eqnarray*}
  R_0 & = & \{\mbox{{\sc if }} (A = A_R) \wedge B_R \mbox{ {\sc then} } A:=\bar{A} \mbox{ {\sc endif}}\},
  \\
  R_3 & = &  \{\mbox{{\sc if }} A \neq A_L \mbox{ {\sc then} } A:=\bar{A} \mbox{ {\sc endif}}\}
  \\
  R_1 = R_2 & = & \{\mbox{{\sc if }} A \neq A_L \mbox{ {\sc then} } A:=\bar{A}, B:=false \mbox{ {\sc endif}},
         \\ & & 
\;\;\mbox{{\sc if }} (A = A_R) \wedge B_R \mbox{ {\sc then} } B:=true \mbox{ {\sc endif}}\}
\end{eqnarray*}
Note that every machine needs at most two different rules in this solution.
We will not argue here about the correctness of this solution, which
was obtained using the synthesis methodology described in the following section and later verified in SAL.
(see~\cite{supporting-sal-models} for the corresponding SAL model).

\section{A Synthesis approach for $FG$ properties}\label{section:synthesisFG}

\newcommand{\xx}{\vec{\mathtt{x}}}
\newcommand{\yy}{\vec{\mathtt{y}}}
\newcommand{\zz}{\vec{\mathtt{z}}}

Roughly speaking, any
template-based program synthesis algorithm must
traverse the
space of possible instantiations of a given template and check
if one of 
%these solutions 
them
satisfies the requirement,\ i.e. implements a 
solution to the given problem.
Checking if a synthesized solution satisfies a requirement
is a formal verification problem. Hence,
synthesis can be simply performed as a loop over the formal 
verification tool. Our approach to synthesis is simpler:
we merge the search and verify loop into just one constraint, 
as done in previous works such as~\cite{DBLP:conf/vmcai/BloemKS14, 
DBLP:journals/sttt/FinkbeinerS13, DBLP:conf/fmcad/AlurBJMRSSSTU13,
DBLP:conf/popl/SrivastavaGF10, DBLP:conf/asplos/Solar-LezamaTBSS06, DBLP:conf/pldi/Solar-LezamaRBE05}.

Our approach can be viewed as a generalization of the idea of
bounded model checking to synthesis.
Just as bounded model checking turns a verification problem
into an {\em{existential}} contraint that encodes a 
{\em{weaker}} version of the
verification problem,
we turn synthesis into a {\em{forall-exists}} constraint
that encodes a {\em{weaker}} version of the synthesis problem.
The key step that makes automated synthesis effective 
is the step that defines the weaker version.
A simpler version of the synthesis problem is obtained by
\\
(i) restricting the universe of possible algorithms
that will be searched and
\\
(ii) replacing the verification step by an approximation step.

In our example, restriction (i) is achieved by fixing a template
of the solution to be synthetized.
That already restricts the search space for possible 
solutions to a finite (but possibly huge) set.
As another example, 
along with fixing the number of processes to a constant,
one can also fix the signature of the messages exchanged
between processes (see ~\cite{DBLP:conf/nfm/GasconT14}).
We believe that 
the limitations derived from
this kind of restrictions are
harmless from the perspective of a system designer or researcher
trying to gain insight into a problem,
although overcoming them in a general way
is a challenging and important 
problem from the verification 
perspective, even 
specifically for 
distributed systems~\cite{parametric-verification-helmut-veith}.

For restriction (ii), we modify 
the property that the solution has to satisfy.
The modification is done so that the new property is easier to verify.
Specifically, and analogously to how bounded model checking
is used to check 
that a property is not violated in a fixed number of steps $k$,
we replace our LTL 
property $FG\phi$ by $X^cG\phi$, for some natural number
$c$.
Intuitively, that corresponds to 
relaxing a property of the form ``eventually, it is always the
case that $\phi$ holds'' to
``after $c$ steps, it is always the
case that $\phi$ holds''. The word ``steps'' might create some
confusion here since it depends on the 
particular problem being analyzed. However, 
for distributed systems, regardless of their
timing model, a notion of step always exist.
Moreover, as we will see in our example, 
an adequate modeling of the problem
might, in some cases,
make the properties
$FG\phi$ and $X^cG\phi$ equivalent for a suitable $c$.

However, the modified property $X^cG\phi$ may still be too complex for our 
synthesis purposes.  Hence, we can replace $G\phi$ 
by just $\phi$ or $\phi\wedge X\phi$.
Thus, instead of meeting the requirement $FG\phi$,
the synthesis tool may find a solution that satisfies
a weaker requirement, say $XXX(\phi)$.
For example, in our running example, 
one possible relaxed version of property (a)
could be:
 At least one rule will always be enabled and
    the system is guaranteed to reach a legitimate state {\em regardless
    of its initial conditions} in $8$ steps.
Whether this property is enough to synthetize a solution 
depends on the provided template.
Similarly, in our previous work on synthesis of distributed consensus
algorithms, our relaxed synthesis property
was that consensus must be achieved in at most $3$ steps,
instead of an arbitrarily large (but finite) number of steps.

Due to the modification of the requirement,
a solution found automatically may not be sound with respect to
the original requirements.  It needs to be formally
verified and hence, % although we relax our $FG$ property
% to make the synthesis problem tractable, 
the synthesized solution is verified against the original property $FG\phi$.
Since our approach leverages existing verification techniques
to the synthesis setting, the final verification step does not need any extra 
encoding or translation work.

\subsection{From the SAL model to the synthesis constraint}

The modeling language of verification tools, such as SAL and NuSMV,
just defines state transition systems, but provide powerful 
language constructs for this purpose that make it easy to model
concurrent systems.
Distributed algorithms, regardless of their timing model, 
can also be easily modeled as open (finite) state transition systems
in these languages.
Let $\xx$ denote all the state variables appearing in a model.
Let $I(\xx)$ be the predicate denoting the initial states
and $T(\xx_1,\xx_2)$ be the predicate denoting the transition relation
(of the state transition system).

SAT-based (bounded) model checking is a powerful
bug-detection technique that is 
available in many verification tools.
Let us provide some details about bounded model checking.
%We denote such set of {\em control} variables
%as $\zz$.
Given the transition system defined by $I$,$T$,
the property $G\phi$,
and a depth to search $3$,
%Also, let $F\phi$ be the requirement, where
%$\phi(\xx)$ is the formula that says that
%$\xx$ is the desired (final) state.
%In bounded model checking, there are no synthesis variables
%and hence the set $\zz$ is empty.
a bounded model checker generates the
following formula:
\begin{eqnarray}
  &&
 \exists{\xx_0,\xx_1,\xx_2,\xx_3}: 
   %\nonumber \\ && \quad
  I(\xx_0) \wedge T(\xx_0,\xx_1) \wedge T(\xx_1,\xx_2) \wedge T(\xx_2,\xx_3) \wedge \neg\phi(\xx_3)
  \label{eq-vc}
\end{eqnarray}
which states that there is $3$-step execution of the
system that violates the property $G\phi$.

Now consider the problem of synthesizing a transition system
to satisfy $F\phi$.
Let $\zz$ denote all the state variables appearing in a
{\em template model/sketch} of the transition system.
The set $\zz$ can be partitioned  as $\xx\cup\yy$, where
the $\yy$ are the (input) variables used to 
represent the synthesis search space 
and $\xx$ are the remaining regular (non-synthesis) variables 
(as in the verification case above).

Instead of synthesizing for $F\phi$,\ i.e. enforcing the LTL property $F\phi$
in the resulting synthetized model, say we decide to 
satisfy the stronger requirement $XXX\phi$.
Given the template transition system defined by $I,T$
with synthesis variables $\yy$ and non-synthesis variables $\xx$,
the property $F\phi$,
and a depth for synthesis $3$,
a {\em{bounded model synthesizer}} generates the following formula: 

\begin{eqnarray}
  &&
 \forall{\yy_0,\yy_1,\yy_2,\yy_3}: \yy_0=\yy_1=\yy_2=\yy_3 \Rightarrow
 \nonumber
 \\ && \quad
 (\exists{\xx_0,\xx_1,\xx_2,\xx_3}:
  I(\zz_0) \wedge T(\zz_0,\zz_1) \wedge T(\zz_1,\zz_2) \wedge T(\zz_2,\zz_3) \wedge \neg\phi(\zz_3))
  \label{eq-sc}
\end{eqnarray}
where $\zz_i = \xx_i\cup\yy_i$ for all $i$.

Formula~\ref{eq-sc} says that ``for every concrete
instance of the state transition system (defined by assignment
to $\yy_0$), there is an execution of that transition system
that does not reach $\phi$ in $3$ steps''.
If this formula is invalid, then it means that there
is a concrete instantiation of the template that always reaches
$\phi$ in $3$ steps.  This indicates that synthesis is successful
(for the requirement $XXX\phi$, and consequently for $F\phi$).
If the formula is valid, then it means that synthesis fails
for the requirement $XXX\phi$.
It is important to remark that this approach, as well as bounded model checking,
assumes that the transition system of the modeled 
state machines is total,\ i.e. there are no deadlock states.

If the domains of all variables in 
Formula~\ref{eq-sc} have finite cardinality,
then the formula can be written as a 
quantified ($\forall\exists$) Boolean formula (QBF),
which can be solved using off-the-shelf QBF solvers.
The synthesized algorithm, if it exists, is obtained 
from the refutation of the formula
generated by the QBF solver in form of a Herbrand model, i.e. 
a valuation for variables $\yy_0,\yy_1,\yy_2,\yy_3$.

\begin{figure}
\begin{center}
\includegraphics[scale=.4]{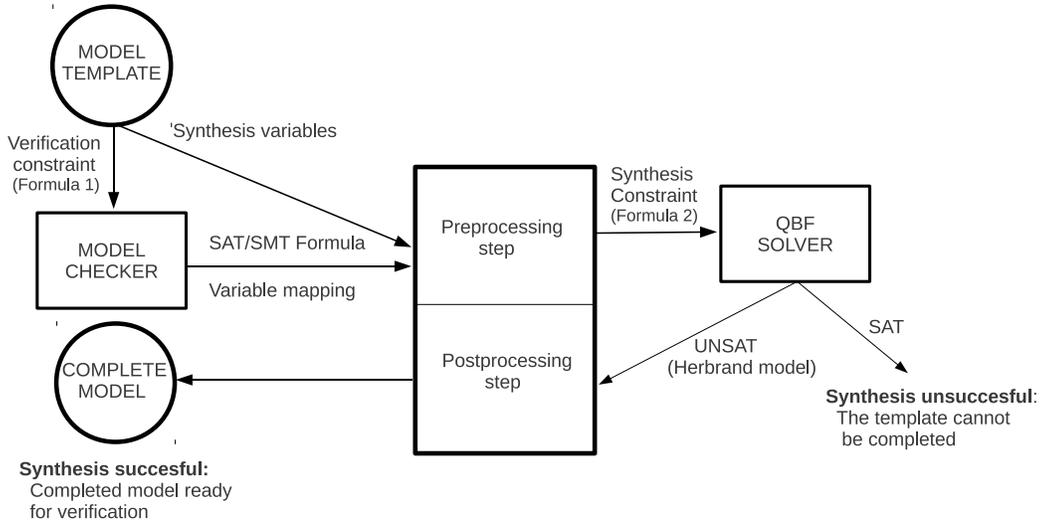}
\end{center}
\caption{The bounded synthesis approach.}
\label{fig:bounded-synthesis-approach}
\end{figure}

Note that Formula~\ref{eq-sc} is not very different from
Formula~\ref{eq-vc}, which is generated by existing
bounded model checkers. 
In the work presented in this paper,
we modeled our template in SAL, and
used the SAL bounded model checker
to generate Formula~\ref{eq-vc},
together with a mapping from variables of the SAL model 
to the corresponding arrays of Boolean variables occurring
in Formula~\ref{eq-vc}.
Then, we used a simple script to convert Formula~\ref{eq-vc} into
Formula~\ref{eq-sc}.  Specifically,
our investigation was carried out
by performing the following steps, 
described also in Figure~\ref{fig:bounded-synthesis-approach}:

\begin{enumerate}\itemsep=0em
  \item We model the {\em template} of distributed algorithm in SAL~\cite{sal-language,DBLP:conf/cav/MouraORRSST04}. The model
    includes synthesis variables $\yy$ to define the transition relation.
  \item We use the SAL bounded model checker to generate the
    SAT formula for the verification constraint (Formula~\ref{eq-vc}).
    The SAT formula implicitly existentially quantifies all variables,
    including the synthesis variables $\yy$.
  \item We modify the SAT formula and convert it into a QBF formula 
    by universally quantifying the synthesis variables.
    (This step uses the mapping from the original
    SAL variables to the Boolean SAT variables).
  \item We use off-the-shelf QBF solvers (and a QBF preprocessor)
    to check validity of the $\forall\exists$ formula (Formula~\ref{eq-sc}).
    For the experiments reported in the next section, we used the QBF preprocessor
    Bloqqer~\cite{bloqqer}, followed by the QBF solver RareQS~\cite{rareqs},
    although we have also experimented with DepQBF~\cite{DBLP:journals/jsat/LonsingB10}.
  \item If the QBF solver returns {\tt{Unsat}}, then the synthesis
    is declared {\em{successful}}, and if the QBF solver returns
    {\tt{Sat}}, then the synthesis process is {\em{unsuccessful}}.
  \item If synthesis is successful, the QBF solver outputs a
    valuation for the synthesis variables $\yy$ (a Herbrand model), which is used to
    obtain a concrete distributed algorithm.
  \item The synthesized algorithm is formally verified:
    if the property was $F\phi$, there is nothing to verify;
    if the property was $FG\phi$, then 
    the property that ``after $k$ steps, the property 
    $\phi$ is always true'' is verified using $k$-induction or symbolic model checking.
\end{enumerate}

\section{Synthesis of a self-stabilizing system with distributed control}\label{dec:running-example}
In this section we present an example of our synthesis approach 
by finding a solution for the problem presented in Section~\ref{sec:running-example}.
Instead of just presenting the template that was provided 
to our synthesis tool to produce a solution, we will ilustrate one 
possible chain of interactions with the synthesis tool that leads to a solution.
Our goal is to demonstrate how interacting with the synthesis tool
is useful to get insight into the problem. To this end, we explain
how Synthia, an imaginary character, used our approach to synthesize
a solution to the problem of Section~\ref{sec:running-example}.
We will not get into the modeling details, 
since all SAL models can be accessed
at~\cite{supporting-sal-models}.
An advantage of our approach is that
limited effort is needed to modify a template
due to the expressivity of the SAL language.

\subsection{How many rules?}
The first question that came to Synthia's mind was whether
a really simple protocol would work. Is a single rule per process enough?
What about the same rule for every process?
That did not seem plausible but, to be sure, Synthia encoded 
the following simple solution template, 
which 
represents a finite family
of possible solutions where 
(1) all $4$ machines have the same rule set,
(2) this rule set contains a single rule,
(3) the privilege of that rule is a 
conjunction of two equality
predicates comparing $A$ and $B$
to two, possibly negated, variables the machine can read.
The corresponding SAL model is {\tt single\_rule.sal} in~\cite{{supporting-sal-models}}.
\[
\begin{array}{l}
R_0 = R_1 = R_2 = R_3 = \{\mbox{{\sc if }}  c_A \wedge c_B \mbox{ {\sc then} } A:=v_A, B:=v_B \mbox{ {\sc endif}}\}, \mbox{ where }
\\ \quad
c_A\in \{(A = b)\mid b\in {\cal D}\}\cup\{(A = \bar{b})\mid b\in {\cal D}\},
\quad {\cal D}= \{A,B,A_L,A_R,B_L,B_R,false,true\},
\\ \quad
c_B\in \{(B = b)\mid b\in {\cal D}\}\cup\{(B = \bar{b})\mid b\in {\cal D}\}, 
 \quad
v_A, v_B\in\{A, \bar{A}, B, \bar{B}, true, false\}.
\end{array}
\]

To confirm her suspicion, Synthia asked the tool whether there is some 
instantiation of this template such that 
the system always reaches a legitimate state in four steps.
Note that the interesting property is in fact $FG(legitimate)$,
which gets transformed into $X^4(legitimate)$.
In about a minute and a half the tool told Synthia that
there is no such instantiation. 
She also tried to synthetize
a solution for $X^8(legitimate)$ and 
$X^{16}(legitimate)$.
As expected, the answer was again ``no'' in 2 and a half and 6 minutes, respectively.

Synthia was convinced that the rules had to implement some way 
of influencing the conditions of the rules of the neighbors. A possibility
is preventing the left neighbor from making a move by having $B_R$ as a condition
of the rule. 
She asked the solver to 
complete the following small variation of the previous template.
The corresponding SAL model is {\tt single\_rule\_BR.sal} in~\cite{{supporting-sal-models}}.
\[
\begin{array}{l}
R_0 = R_1 = R_2 = R_3 = \{\mbox{{\sc if }}  c_A \wedge B_R \mbox{ {\sc then} } A:=v_A, B:=v_B \mbox{ {\sc endif}}\}, \mbox{ where }
\\ \quad
c_A\in \{(A = b)\mid b\in {\cal D}\}\cup\{(A = \bar{b})\mid b\in {\cal D}\},
\quad {\cal D}= \{A,B,A_L,A_R,B_L,B_R,false,true\},\mbox{ and}
\\ \quad
v_A, v_B\in\{A, \bar{A}, B, \bar{B}, true, false\}.
\end{array}
\]

After getting a negative answer for both $X^4(legitimate)$ 
and $X^{16}(legitimate)$ in less than $5$ seconds,
Synthia realized that the symmetry of the rules has to be broken somehow
since otherwise the states where $\forall i\in\{0,\ldots,3\}: B_i = false$ would 
not have a successor. A possibility is to fix 
the value of $B$ in machine $3$ to $true$.
She tried that and, additionally,
fixing the 
value of $B$
in machine $0$ to value $false$,
getting a negative answer in both cases.

To gain more intuition into the problem, Synthia
tried to synthetize a solution
assuming a particular initial state, changing the previous template to obtain the following 
(the corresponding SAL model is {\tt single\_rule\_B\_blocks\_initialized.sal} in~\cite{supporting-sal-models}):
\[
\begin{array}{l}
B_0\mbox{ initialized to }1\\
A_0, A_1, A_2, A_3, B_1, B_2, B_3\mbox{ initialized to }0\\
R_0 = R_1 = R_2 = R_3 = \{\mbox{{\sc if }}  c_A \wedge B_R \mbox{ {\sc then} } A:=v_A, B:=v_B \mbox{ {\sc endif}}\}, \mbox{ where }
\\ \quad
c_A\in \{(A = b)\mid b\in {\cal D}\}\cup\{(A = \bar{b})\mid b\in {\cal D}\},
\quad {\cal D}= \{A,B,A_L,A_R,B_L,B_R,false,true\},
\\ \quad
v_A, v_B\in\{A, \bar{A}, B, \bar{B}, true, false\}.
\end{array}
\]

The enforced property was again $X^4(legitimate)$.
The answer of the solver, in less than $2$ seconds, was 

\[
\begin{array}{l}
B_0\mbox{ initialized to }1\\
A_0, A_1, A_2, A_3, B_1, B_2, B_3\mbox{ initialized to }0\\
R_0 = R_1 = R_2 = R_3 = \{\mbox{{\sc if }}  B_R \mbox{ {\sc then} } A:=B_R, B:=A \mbox{ {\sc endif}}\}
\end{array}
\]

Synthia knew that this could not be generalized, since
her former attempt to synthetize a solution had failed.
She tried to verify the previous solution for the property
$X^4(legitimate)$ using symbolic model checking. It worked.
The next step then, is to test $FG(legitimate)$. SAL returned 
a counterexample of length $10$. Also, simulating 
by hand the execution of the previous complete model helped
Synthia to get convinced that a
solution where $R_0 = R_1 = R_2 = R_3$
could not exist, although she did not worry about formally proving it.

\subsection{Two rules per machine}
Synthia extended the template to have two rules per machine.
The corresponding SAL model is {\tt two\_rules\_general.sal} in~\cite{supporting-sal-models}.
\[
\begin{array}{l}
R_i = \{\mbox{{\sc if }} c_{A,i,1} \wedge c_{B,i,1} \mbox{ {\sc then} } A:=v_{A,i,1}, B:=v_{B,i,1} \mbox{ {\sc endif}},
\\ \quad\quad~~\mbox{{\sc if }} c_{A,i,2} \wedge c_{B,i,2} \mbox{ {\sc then} } A:=v_{A,i,2}, B:=v_{B,i,2} \mbox{ {\sc endif}} \}, \mbox{ where }
\\[0.7ex] \quad
c_{A,i,j}\in \{(A = b)\mid b\in {\cal D}\}\cup\{(A = \bar{b})\mid b\in {\cal D}\},
\quad {\cal D}= \{A,B,A_L,A_R,B_L,B_R,false,true\},
\\ \quad
c_{B,i,j}\in \{(B = b)\mid b\in {\cal D}\}\cup\{(B = \bar{b})\mid b\in {\cal D}\}, 
 \quad
v_{A,i,j}, v_{B,i,j}\in\{A, \bar{A}, B, \bar{B}, true, false\}\\[0.7ex]
\quad\quad~~\mbox{for every } i\in\{0,1,2,3\}, j\in\{1,2\}.
\end{array}
\]
Once again, the enforced property was $X^4(legitimate)$, the bounded version of
$FG(legitimate)$.  The solver did not produce a solution in $10$ minutes and 
Synthia lost her patience. It is important to remark here that 
the election for the value of the constant $c$ in the strengtening $X^c(\phi)$ of 
$FG(\phi)$ may be crucial to obtain a solution.
With this fact in mind, Synthia tried $X^{12}(legitimate)$ with no success.
The 2QBF $\forall\exists$ instance corresponding to formula~\ref{eq-sc}  
for this system template and the property $X^4(legitimate)$
has $128$ universal and $23273$ existential variables.
The QBF preprocessor Bloqqer~\cite{bloqqer} reduced the number of clauses from 
$91714$ to $15338$.

Synthia knew that this template was too general, and thus many of its instances are 
either equivalent to some other instance or can be trivially discarded.
The goal then, was to find a more restrictive template and reduce
the number of universal variables in the resulting 2QBF problem.
A simple option is considering the restriction of the previous template where
$R_1 = R_2$. This requires trivial changes with respect to the previous
template and is encoded in {\tt two\_rules\_reduced.sal}.
The resulting QBF formula for the property $X^4(legitimate)$, 
after being preprocessed with bloqqer, 
has $96$ universal variables, 
$19059$ existential variables, and $16099$ clauses.
As before, Synthia gives up after waiting for around $15$ minutes.

After realizing that the tool will not give her all the answers,
Synthia decided to go back to the idea of using $B_R$
to block the left neighbor from making a move.
Note that that case is not covered in the previous template.
Additionally, she kept the restriction that $R_1 = R_2$.
The resulting SAL model is {\tt two\_rules\_reduced\_BR.sal}.

\noindent
$B_1$ is fixed to have value $false$,
and the set of rules for each of the four machines is defined as follows:
\[
\begin{array}{l}
R_i = \{\mbox{{\sc if }} c_{A,k,1} \wedge B_R \mbox{ {\sc then} } A:=v_{A,k,1}, B:=v_{B,k,1} \mbox{ {\sc endif}},\\ \quad\quad~~\mbox{{\sc if }} c_{A,k,2} \wedge c_{B,k,2} \mbox{ {\sc then} } A:=v_{A,k,2}, B:=v_{B,k,2} \mbox{ {\sc endif}} \}, \mbox{ where }
\\[0.7ex] \quad
 k = 2\mbox{ if }i = 3\mbox{ and }k = i\mbox{, otherwise and}
\\ \quad
c_{A,k,j}\in \{(A = b)\mid b\in {\cal D}\}\cup\{(A = \bar{b})\mid b\in {\cal D}\},
\quad {\cal D}= \{A,B,A_L,A_R,B_L,B_R,false,true\},
\\ \quad
c_{B,k,2}\in \{(B = b)\mid b\in {\cal D}\}\cup\{(B = \bar{b})\mid b\in {\cal D}\}, 
 \quad
v_{A,k,j}, v_{B,k,j}\in\{A, \bar{A}, B, \bar{B}, true, false\}\\[0.7ex]
\quad\quad~~\mbox{for every } i\in\{0,1,2,3\}, j\in\{1,2\}.
\end{array}
\]

After not obtaining a solution from the QBF solver in $10$ minutes,
Synthia decided to simplify her template even more, 
by restricting the domains of the conditions and the assignments of the
rules as $c_{A,k,j}\in \{(A = b)\mid b\in D\}\cup\{(A = \bar{b})\mid b\in {\cal D}\}$,
where ${\cal D}= \{B,A_L,A_R\}$, and 
$v_{A,k,j}, v_{B,k,j}\in\{A, \bar{A}, true, false\}$.
Again, the enforced property is $X^4(legitimate)$.
In ~$5$ minutes the tool reported that 
there was no instance of the template satisfying the property.

Synthia was confused at this point.
After using the tool to synthesize a solution for a particular case, 
Synthia realized that, not only the value of $B$ in machine $0$, 
but also the value of $B$ in machine $3$, must be fixed.
The corresponding SAL model is {\tt two\_rules\_reduced\_BR\_simpl\_values.sal}.
When enforcing $X^4(legitimate)$, the tool found an instance
not satisfying $FG(legitimate)$, which was easily detected
when trying to formally verify it.
For the case of
$X^{12}(legitimate)$, the tool found the following solution:\\[2ex]
\noindent
$B_1$ is fixed to have value $false$, $B_4$ is fixed to have value $true$,
and the set of rules for each of the four machines is defined as follows:
\begin{eqnarray*}
  R_0 & = & \{\mbox{{\sc if }} (A = B) \wedge B_R \mbox{ {\sc then} } A:=\bar{A} \mbox{ {\sc endif}},
      \\ & & 
\;\;\mbox{{\sc if }} (A = B) \mbox{ {\sc then} } A:=\bar{A} \mbox{ {\sc endif}}\}\},
  \\
R_3 & = &  \{\mbox{{\sc if }} A \neq A_L \wedge B = A_R \mbox{ {\sc then} } A:=\bar{A} \mbox{ {\sc endif}}\}
  \\
  R_1 = R_2 & = & \{\mbox{{\sc if }} A \neq A_r \wedge B_R \mbox{ {\sc then} } A:=\bar{A}, B:=false \mbox{ {\sc endif}}\}
\end{eqnarray*}

The first thing that Synthia did was verifying
$FG(legitimate)$ to make sure that the synthetized 
solution preserves stabilization. This property
could be proved by SAL using symbolic model checking
in a few seconds. Also, the solution was checked for deadlock states.
 %Also, k-induction could prove
%$X(X(X(F())))$ giving us a bound for the number of steps
%needed to reach stabilization.
However, after inspecting the solution, Synthia realized that
it is not fair. It was confirmed
by trying to verify the more complex
property $FG(legitimate\wedge M)$, where $M$ is a predicate
that is satisfied iff every machine made a move at some step in the past,
since symbolic model checking produced a counterexample.

Recall that the notion of fairness required
in our example is that, for every pair of machines $m_i, m_j$, 
there is a sequence of steps of the system going from every legitimate state
where a rule of $m_i$ is enabled to a legitimate state
were a rule of $m_j$ is enabled.
Note that this property requires the existence of an execution,
and hence it intuitively corresponds to the $E$ (Exists) temporal operator
in Computational Tree Logic (CTL), and not the 
$F$ operator in LTL. 
In the original problem presented by Dijkstra,
the definition of enabled rule did not require
a rule to change the current state to 
be enabled. However, note that
every execution of the system
with Dijkstra's definition of enabled can be 
associated to an execution in our setting.
Hence,
the property $FG(legitimate\wedge M)$
correctly captures the original fairness condition.

Hence, Synthia used our tool to synthesize a solution for 
$X^{12}(legitimate\wedge M)$, and obtained, in less than $30$ seconds,
the solution presented in Section~\ref{sec:running-example}.

The first question that came to Synthia's mind was whether
the synthesized solution could be generalized to $n$ machines.
However, before getting into that, Synthia asked one last question to the synthesis tool:
is there any instantiation of the template satisfying $X^{11}(legitimate)$?
The tool quickly answered ``no''. Synthia started wondering whether
that bound holds for any algorithm satisfying the requirements.
She then closed her laptop and grabbed pencil and paper.

\section{Conclusion and further work}~\label{sec:discussion}
We have presented a practical approach to the synthesis
of finite-state distributed systems based in bounded synthesis of LTL properties.
Our approach can be seen as a natural first step in the extension of
the capabilities of a model checker to synthesis and builds up
on the fact that, while synthetizing a complex system from
scratch is still unfeasible in practice, the  
recent progress in QBF solving enables synthesis from 
human-provided templates.

As further work, we plan to extend an existing model checker such as SAL
to have synthesis capabilities. 
While the SAL language is very appropiate for the modeling of 
distributed systems, it does not provide
specific constructs for describing templates.
An important component of this task is
the design and implementation of an extension of 
the SAL language to support definition of templates.

From another perspective, besides experimenting more with our approach,
we are interested in leveraging it to the $k$-induction and
infinite settings. The latter is enabled by the recent progress in
$\exists\forall$ SMT solving. However, 
more investigation is needed in finding decision procedures for that 
problem that are well suited for the instances that have to be solved in our setting.

%We should mention the characteristics of the QBF problems in every example.
%Maybe a table for all of them? We should make them available too.

%Turns out that Synthia could have found a fair solution in less than 
%two minutes by fixing the value of $B$ in machines $0$ and $3$.
%That problem has $68$ boolean synthesis variables and $41143$ exstential boolean variables. 

%It might be worth trying to synthetize a variant of the
%first template with two rules that allows to choose 
%left hand sides of conditions.
%That plus fixing B for m0 and m3 can be instantiated to a solution.
%This is more general than Synthia's initial attempt.
%I started doing it in two\_rules\_more\_general.
\bibliographystyle{eptcs}
\bibliography{main}
\end{document}